\documentclass[aps,nofootinbib,preprintnumbers,showpacs,prd,
twocolumn,superscriptaddress]{revtex4-2}
\usepackage{array}
\usepackage{booktabs}
\usepackage{subfigure}
\usepackage{capt-of}
\usepackage{graphicx}
\usepackage{lipsum}
\usepackage{epstopdf}
\usepackage{amsmath}
\usepackage{bm}
\usepackage{amssymb}
\usepackage{color,xcolor}
\usepackage[bookmarks=false]{hyperref}
\hypersetup{colorlinks=true,
            citecolor=green,
            linkcolor=blue,
            urlcolor=blue,
            pdfstartview=FitH,
            bookmarksopen=true}

\begin{document}
	
\title{Higher-twist generalized parton distributions of the pion and kaon at zero skewness in the light-cone quark model}
	
\author{Xiaoyan Luan}
\author{Zhun Lu}\email[]{zhunlu@seu.edu.cn}
\affiliation{School of Physics, Southeast University, Nanjing 211189, China}
	
	\begin{abstract}
    We investigate the higher-twist generalized parton distributions (GPDs) of the pion and kaon at zero skewness by adopting the overlap representation within the light-cone formalism. Using the wave functions of pion and kaon deduced from a light-cone quark model (LCQM), we calculate the twist-3 and twist-4 GPDs $E_{2}(x,0,t)$, $G_{2}(x,0,t)$, $F_{3}(x,0,t)$ and $H_{3}(x,0,t)$ of valence quark inside the pion and kaon mesons. Numerical results for these higher-twist GPDs are presented. By taking the forward limit, we also present the numerical results for the corresponding twist-3 and twist-4 parton distribution functions (PDFs) $e(x)$ and $f_{3}(x)$ of pion and kaon. We further study the relations between $e(x)$ or $f_{3}(x)$ and the twist-2 unpolarized PDF $f_{1}(x)$.
	\end{abstract}
	\maketitle
	
\section{Introduction}\label{Sec1}

	Understanding the internal structure of hadrons composed of constituent quarks, gluons and sea quarks is one of the main goals of QCD and hadronic physics. 
The generalized parton distributions (GPDs)~\cite{Muller:1994ses,Ji:1996nm,Radyushkin:1997ki,Diehl:2015uka}, usually viewed as the extension of the standard parton distribution functions (PDFs), have been recognized as important quantities describing the three-dimensional structure of the nucleon in addition to the transverse momentum dependent parton distributions (TMDs). 
The GPDs not only depend on the longitudinal momentum faction $x$ of the parton, but also can probe the square of the total momentum transferred $t$ and the longitudinal momentum transferred skewness $\xi$.
As the off-forward matrix elements of nonlocal operators, the GPDs are experimentally accessible through the deeply virtual Compton scattering (DVCS)~\cite{Ji:1996nm,Radyushkin:1996nd,Belitsky:2001ns} or the deeply virtual meson production (DVMP)~\cite{Goloskokov:2006hr,Goloskokov:2007nt,Goloskokov:2009ia,Goloskokov:2011rd}. 
The experimental data of these processes were collected by the H1 collaboration~\cite{H1:1999pji,H1:2001nez,H1:2005gdw}, the ZEUS collaboration \cite{ZEUS:1998xpo,ZEUS:2003pwh}, as well as the fixed target experiments at HERMES \cite{HERMES:2001bob,HERMES:2011bou,HERMES:2012gbh}, COMPASS~\cite{dHose:2004usi}, and JLab~\cite{CLAS:2001wjj}. 
On the one hand, the GPDs can reduce to the standard PDFs in the forward limit which indicate the GPDs encode more information about the partonic structure of the hadron than one-dimensional PDFs. 
On the other hand, the GPDs can also be connected with the form factors (FFs) and orbital angular momentum (OAM) of the quarks and gluons inside the nucleon~\cite{Radyushkin:1997ki,Ji:1996nm,Sehgal:1974rz,Kroll:2020jat}. 
Through a Fourier transform with respect to the transverse momentum transfer $\Delta_T$, one can obtain the distributions in the impact parameter space which provide tomographic description of the nucleon structure.
	
Distribution functions are classified according to their twist, indicating the order of the $1/Q$ expansions in which they appear, where $Q$ represents the hard scale of the physical process. Most studies related to GPDs focus on the leading-twist, while the knowledge of the higher-twist GPDs is still limited. 
However, higher-twist GPDs give access to a wealth of information about the partonic structure of hadrons. There are several motivations for measuring higher-twist GPDs, specifically the twist-3 ones. 
Firstly, the twist-3 effects could be measured at relatively low $Q^2$, which indicate the twist-3 GPDs are not negligible for the energy scales explored experimentally. 
Secondly, another motivation for exploring twist-3 GPDs is the relation between them and the OAM of quarks inside a longitudinally polarized nucleon~\cite{Ji:1996ek, Penttinen:2000dg}. 
Twist-3 GPDs have also been shown to provide information about the average transverse color Lorentz force acting on quarks~\cite{Burkardt:2008ps, Aslan:2019jis}. Furthermore, certain spin-orbit correlations of the nucleon and the twist-3 DVCS amplitude can be expressed through twist-3 GPDs~\cite{Lorce:2014mxa,Bhoonah:2017olu,Guo:2022cgq}. Finally, higher-twist GPDs can be related to generalized transverse momentum dependent parton distributions (GTMDs)~\cite{Meissner:2008ay,Meissner:2009ww,Lorce:2013pza,Rajan:2017cpx}, the so-called ``mother distributions" which contain the most complete structural information of hadrons.
	
Despite their importance, higher-twist distributions are difficult to determine experimentally because of the challenge of isolating them from the leading-twist contributions. The contributions of the higher-twist are usually smaller than that of the leading-twist due to the existence of suppression factors. 
In recent years, the kinematical twist-3 effects~\cite{Anikin:2000em,Penttinen:2000dg,Belitsky:2000vx,Kivel:2000cn,
Radyushkin:2000ap,Guo:2021gru} and twist-4 effects such as finite $t$ and target mass corrections~\cite{Blumlein:2006ia,Blumlein:2009hit,Braun:2011zr,Braun:2011dg,Braun:2012bg,Braun:2012hq,Braun:2014sta} in the DVCS process have been studied extensively. Moreover, the higher-twist distributions have also been investigated in various models. 
For example, calculation on higher-twist TMDs has been performed by the MIT bag model~\cite{Signal:1996ct,Avakian:2010br,Lorce:2014hxa}, the diquark spectator models~\cite{Lu:2012gu,Mao:2013waa,Mao:2014aoa,Liu:2021ype}, and models based on the light-front approaches~\cite{Burkardt:2001iy,Kundu:2001pk,Mukherjee:2009uy,
Lorce:2016ugb,Pasquini:2018oyz,Sharma:2023azu,Sharma:2023wha,Puhan:2023ekt,Zhu:2023lst}. Other models such as the chiral quark soliton models~\cite{Schweitzer:2003uy,Wakamatsu:2003uu,Wakamatsu:2007nc,Ohnishi:2003mf,
Cebulla:2007ej}, the instanton model~\cite{Balla:1997hf,Dressler:1999hc} have also been applied to calculate higher-twist distributions. 
Compared to higher-twist TMDs, the model calculations for GPDs at higher-twist are seldom. 
Meanwhile, the twist-3 GPDs have been investigated in the quark target model~\cite{Mukherjee:2002pq,Mukherjee:2002xi,Aslan:2018tff} and scalar diquark model~\cite{Aslan:2018tff} for the nucleon. 
The twist-4 proton GTMDs were studied within the light-front quark-diquark model (LFQDM) recently~\cite{Sharma:2023tre},
and the chiral-even twist-3 GPDs of the proton have also been investigated by the Lattice QCD~\cite{Bhattacharya:2023nmv}. 
In two recent works, the calculations of the twist-3 GPDs or the twist-4 chiral-even GPDs of the proton have been carried out in the basis light-front quantization (BLFQ)~\cite{Zhang:2023xfe} and LFQDM~\cite{Sharma:2023ibp} with the overlaps of the LFWFs, respectively. 
In addition, for spin-0 target, the parameterization of quark GPDs up to twist-4 has been given in Ref.~\cite{Meissner:2008ay}.

In this paper, we apply the light-cone quark model (LCQM) to calculate the twist-3 and twist-4 GPDs of the valence quark inside the pion and kaon at zero skewness. 
As pointed out in Ref.~\cite{Diehl:2000xz,Brodsky:2000xy}, the light-cone formalism is a suitable approach for studying GPDs. 
Similar to the twist-2 GPDs, the higher-twist GPDs can be also expressed as the overlap integration of hadronic light-cone wave functions (LCWFs). 
The minimal LCWFs of the spin-0 pseudoscalar mesons can be expressed in terms of the $|q\bar{q} \rangle$ Fock states, as calculated in Ref.~\cite{Xiao:2003wf,Qian:2008px}. 
Thus, the expressions of the twist-3 and twist-4 GPDs of spin-0 pseudoscalar mesons ($E_{2}(x,0,t)$, $G_{2}(x,0,t)$, $F_{3}(x,0,t)$ and $H_{3}(x,0,t)$) can be obtained within the overlap representation in terms of these LCWFs. 
The numerical results for these higher-twist GPDs are presented by properly choosing the values of the parameters. 
By taking the forward limit, the numerical results for the corresponding twist-3 and twist-4 PDFs of the pion and kaon are also presented. 
Furthermore, the relations between twist-3 and twist-4 PDFs and the twist-2 unpolarized PDF are analyzed. 
As a check, we compare our numerical results for higher-twist PDFs with previous model calculations~\cite{Lorce:2016ugb,Kaur:2020vkq,Puhan:2023ekt,Zhu:2023lst}.
	
The remaining content of the work is as follows. 
In Sec. II, we briefly introduce the LCQM and present the LCWFs of  pion and kaon mesons. We introduce the definition of the higher-twist GPDs in Sec. III, where we also derive the analytic expressions of twist-3 and twist-4 GPDs of the valence quark for the pion and kaon using the overlap representation in terms of LCWFs. 
In Sec. IV, we present the numerical results for the twist-3 and twist-4 GPDs and PDFs. 
We summarize this work in Sec. V.
	
\section{Light-cone wave functions of the Pion and kaon }\label{Sec2}
	
For the spin-0 pseudoscalar mesons such as the pion and the kaon, the Fock states can be cast into
\begin{align}
		|\pi(K)\rangle & = |q \bar{q}\rangle+|q \bar{q} g\rangle+\cdots .
\end{align}
The light-cone Fock-state expansion of the corresponding hadronic eigenstate $\left|M\left(P^{+}, \boldsymbol{P}_{T}, S_{z}\right)\right\rangle$ in terms of its constituent eigenstates $\left|n \right\rangle$ is given by
	\begin{align}
		\notag	\left|M\left(P^{+}, \boldsymbol{P}_{T}, S_{z}\right)\right\rangle & = \sum_{n} \int \prod_{i = 1}^{n} \frac{\mathrm{d} x_{i} \mathrm{~d}^{2} k_{T i}}{\sqrt{x_{i}} 16 \pi^{3}} 16 \pi^{3} \delta\left(1-\sum_{i = 1}^{n} x_{i}\right) \\\notag
		& \times \delta^{(2)}\left(\sum_{i = 1}^{n} \boldsymbol{k}_{T i}\right) \psi_{n/M}\left(x_{i}, \boldsymbol{k}_{T i}, \lambda_{i}\right) \\
		& \times\left|n ; x_{i} P^{+}, x_{i} \boldsymbol{P}_{T}+\boldsymbol{k}_{T i}, \lambda_{i}\right\rangle,
	\end{align}
where $P = (P^{+}, P^{-},\boldsymbol{P}_{T})$ is the four-vector momentum of the meson in the light-cone coordinate, $S_{z}$ is the longitudinal spin projection of the target and $\lambda_{i}$ denotes the helicity of parton $i$. 
In the follow ing calculation, we only consider the first order contribution which is the minimal Fock-state of the pion/kaon meson $(n = 2)$. 
In this case, the wave function $\psi_{M}\left(x, \boldsymbol{k}_{T}, \lambda_{q}, \lambda_{\bar{q}}\right)$ corresponds to the two-particle states $\left|x, \boldsymbol{k}_{T}, \lambda_{q}, \lambda_{\bar{q}}\right\rangle$, with $\lambda_{q}$ and $\lambda_{\bar{q}}$ are the helicities of quark and the spectator antiquark.
 
 In the light-cone frame, the momenta of the meson $(P)$, constituent quark $(k_{1})$ and antiquark $(k_{2})$ are defined as
  \begin{align}
 	P & =\left(P^{+}, \frac{M_{\pi(K)}^{2}}{P^{+}}, \boldsymbol{0}_{T}\right), \\
 	k_{1} & =\left(x P^{+}, \frac{\boldsymbol{k}_{T}^{2}+m_{q}^{2}}{x P^{+}}, \boldsymbol{k}_{T}\right), \\
 	k_{2} & =\left((1-x) P^{+}, \frac{\boldsymbol{k}_{T}^{2}+m_{\bar{q}}^{2}}{(1-x) P^{+}},-\boldsymbol{k}_{T}\right).
  \end{align}
 The LCWF in the LCQM is written as
 \begin{align}
 	\psi_{S_{z}}^{F}\left(x, \boldsymbol{k}_{T}, \lambda_{q}, \lambda_{\bar{q}}\right) & = \varphi\left(x, \boldsymbol{k}_{T}\right) \chi_{S_{z}}^{F}\left(x, \boldsymbol{k}_{T}, \lambda_{q}, \lambda_{\bar{q}}\right),
 \end{align}
where $\chi_{S_{z}}^{F}\left(x, \boldsymbol{k}_{T}, \lambda_{q}, \lambda_{\bar{q}}\right)$ is the spin wave function, and $\varphi\left(x, \boldsymbol{k}_{T}\right)$ is the momentum space wave function. 
The superscript $F$ stands for the front form. 
The LCWFs of pion and kaon have been derived in Ref.~\cite{Xiao:2003wf,Qian:2008px}. 
For the LCWFs of pion, we take the following form~\cite{Xiao:2003wf}
 \begin{align}\label{piLCWFs}
		\notag \psi^{\pi}_{0}\left(x, \boldsymbol{k}_{T},\uparrow,\downarrow\right) & = +\frac{m}{\sqrt{2\left(m^{2}+\boldsymbol{k}_{T}^{2}\right)}} \varphi^{\pi} \quad\left(l^{z} = 0\right), \\\notag
		\psi^{\pi}_{0}\left(x, \boldsymbol{k}_{T},\downarrow,\uparrow\right) & = -\frac{m}{\sqrt{2\left(m^{2}+\boldsymbol{k}_{T}^{2}\right)}} \varphi^{\pi} \quad\left(l^{z} = 0\right), \\\notag
		\psi^{\pi}_{0}\left(x, \boldsymbol{k}_{T},\uparrow,\uparrow\right) & = -\frac{k_{1}-i k_{2}}{\sqrt{2\left(m^{2}+\boldsymbol{k}_{T}^{2}\right)}} \varphi^{\pi} \quad\left(l^{z} = -1\right), \\
		\psi^{\pi}_{0}\left(x, \boldsymbol{k}_{T},\downarrow,\downarrow\right) & = -\frac{k_{1}+i k_{2}}{\sqrt{2\left(m^{2}+\boldsymbol{k}_{T}^{2}\right)}} \varphi^{\pi} \quad\left(l^{z} = +1\right),
\end{align}
where $x$ and $\boldsymbol{k}_{T}$ denote the longitudinal momentum fraction and the transverse momentum of the quark inside the pion/kaon, respectively.
Here, we adopt the same value for the masses of the quark and the antiquark mass in the pion: $m_{q}=m_{\bar{q}}=m$.

For the kaon LCWFs, we adopt the following form from Refs.~\cite{Xiao:2003wf,Kaur:2019jow}
\begin{align}\label{KLCWFs}
		\notag \psi^{K}_{0}\left(x, \boldsymbol{k}_{T},\uparrow,\downarrow\right) & = +\frac{(1-x)m_{1}+xm_{2}}{\sqrt{2\left(L^{2}+\boldsymbol{k}_{T}^{2}\right)}} \varphi^{K} \quad\left(l^{z} = 0\right), \\\notag
		\psi^{K}_{0}\left(x, \boldsymbol{k}_{T},\downarrow,\uparrow\right) & = -\frac{(1-x)m_{1}+xm_{2}}{\sqrt{2\left(L^{2}+\boldsymbol{k}_{T}^{2}\right)}} \varphi^{K} \quad\left(l^{z} = 0\right), \\\notag
		\psi^{K}_{0}\left(x, \boldsymbol{k}_{T},\uparrow,\uparrow\right) & = -\frac{k_{1}-i k_{2}}{\sqrt{2\left(L^{2}+\boldsymbol{k}_{T}^{2}\right)}} \varphi^{K} \quad\left(l^{z} = -1\right), \\
		\psi^{K}_{0}\left(x, \boldsymbol{k}_{T},\downarrow,\downarrow\right) & = -\frac{k_{1}+i k_{2}}{\sqrt{2\left(L^{2}+\boldsymbol{k}_{T}^{2}\right)}} \varphi^{K} \quad\left(l^{z} = +1\right), 
\end{align}
with 
\begin{align}
		L^{2} & = (1-x) m_{1}^{2}+x m_{2}^{2}-x(1-x)\left(m_{1}-m_{2}\right)^{2} .
\end{align}
where $m_{q}=m_{1}$ and $m_{\bar{q}}=m_{2}$ of which the values are different for the kaon.  
These terms of the spectator antiquark can be easily determined as $1-x$ and $-\boldsymbol{k}_{T}$. We use $\uparrow(\downarrow)$ to denote the positive (negative) helicity of the struck quark and the spectator antiquark.

For the wave function in momentum space $\varphi_{M}$, we adopt the Brodsky-Huang-Lepage (BHL) prescription~\cite{Xiao:2003wf,Qian:2008px}
\begin{align}
\varphi^{\pi}\left(x, \boldsymbol{k}_{T}\right) & = A^{\pi} \exp \left[-\frac{1}{8 \beta_{\pi}^{2}} \frac{\boldsymbol{k}_{T}^{2}+m^{2}}{x(1-x)}\right],\\
		\notag\varphi^{K}\left(x, \boldsymbol{k}_{T}\right)& = A^{K} \exp \left[-\frac{\frac{\boldsymbol{k}_{T}^{2}+m_{1}^{2}}{x}
+\frac{\boldsymbol{k}_{T}^{2}+m_{2}^{2}}{1-x}}{8 \beta_{K}^{2}}\right.\nonumber\\
&\left.-\frac{\left(m_{1}^{2}-m_{2}^{2}\right)^{2}}{8 \beta_{K}^{2}\left(\frac{\boldsymbol{k}_{T}^{2}+m_{1}^{2}}{x}
+\frac{\boldsymbol{k}_{T}^{2}+m_{2}^{2}}{1-x}\right)}\right],
\end{align}
where $A^{\pi}$ and $A^{K}$ are the normalization constants for the pion and kaon, respectively.

\section{Twist-3 and Twist-4 GPDs in the overlap representation}\label{Sec3}
	
In this section, we present the calculations for the twist-3 and twist-4 GPDs of the pion and kaon at zero skewness in the LCQM using the overlap representation. The light-cone formalism has been widely applied in the calculation of PDFs of nucleons and mesons~\cite{Lepage:1980fj}. 
Within the light-cone approach, a hadronic composite state can be expressed as LCWFs on the Fock-state basis. 
The overlap representation has also been used to study various FFs of the hadrons~\cite{Brodsky:2000ii,Xiao:2003wf}, anomalous magnetic moment of the nucleon~\cite{Brodsky:2000ii}, TMDs~\cite{Bacchetta:2008af,Lu:2006kt,Luan:2022fjc}, GPDs~\cite{Muller:2014tqa,Brodsky:2000xy,Burkardt:2003je,Luan:2023lmt} as well as the quark Wigner distributions~\cite{Ma:2018ysi,Kaur:2020par,Luan:2024nwc}. Here we extend light-cone formalism to calculate the higher-twist GPDs of the spin-0 pseudoscalar mesons. 
	
For spin-0 pseudoscalar mesons, the valence quark GPDs are defined from the quark-quark correlation function, which is written as~\cite{Meissner:2008ay}
	\begin{align}
		\notag&F^{[\Gamma]}(x, \xi, t) = \int \frac{d z^{-}}{2 \pi}e^{ik \cdot z}\\
		&\left. \times\left\langle p^{\prime}\left|\bar{\psi}\left(-\frac{z}{2}\right) \Gamma\mathcal{W}\left(-\frac{z}{2}, \frac{z}{2}\right)  \psi\left(\frac{z}{2}\right)\right| p \right\rangle\right|_{z^{+} = 0, \vec{z}_{T} = 0},
	\end{align}
Here, $\mathcal{W}$ is the gauge link connecting the quark fields at positions $-\frac{z}{2}$ and $\frac{z}{2}$ to ensure the color gauge invariance, $P=(p+p^{\prime})/2$ is the average four-momentum of the nucleon, $x=k^+/P^+$ is the average fraction of the light-cone momentum carried by the active quark, $\Delta=p^{\prime}-p$ is the momentum transfer to the proton, and $t=\Delta^{2}=-{\Delta}_T^2$ represents the momentum transfer squared. In addition, $\xi=-\Delta^{+}/2 P^{+}$ is the skewness parameter. 
In this work, we take the twist-3 and the twist-4 Dirac matrices, i.e., $\Gamma=\{1,\gamma^{j}, \gamma^{j} \gamma^{5}, i \sigma^{i j} \gamma_{5}$, or $\gamma^{-},i \sigma^{j-} \gamma_{5}\}$.

According to the parameterization in Ref.~\cite{Meissner:2008ay}, the nonvanishing GPDs at twist-3 and two GPDs at twist-4 are defined as
	\begin{align}
		F^{[1]} & = \frac{M}{P^{+}}\left[E_{2}(x, \xi, t)\right], \\
		F^{\left[\gamma^{j}\right]} & = \frac{M}{P^{+}}\left[\frac{\Delta_{T}^{j}}{M} F_{2}(x, \xi, t)\right],\\
		F^{\left[\gamma^{j} \gamma_{5}\right]} & = \frac{M}{P^{+}}\left[\frac{i \varepsilon_{T}^{i j} \Delta_{T}^{i}}{M} G_{2}(x, \xi, t)\right], \\
		F^{\left[i \sigma^{i j} \gamma_{5}\right]} & = \frac{M}{P^{+}}\left[i \varepsilon_{T}^{i j} H_{2}(x, \xi, t)\right], \\
		F^{\left[\gamma^{-}\right]} & = \frac{M^{2}}{\left(P^{+}\right)^{2}}\left[F_{3}(x, \xi, t)\right], \\
		F^{\left[i \sigma^{j-} \gamma_{5}\right]} & = \frac{M^{2}}{\left(P^{+}\right)^{2}}\left[\frac{i \varepsilon_{T}^{i j} \Delta_{T}^{i}}{M} H_{3}(x, \xi, t)\right].
	\end{align} 
where $\sigma^{i j}=i\left[\gamma^{i}, \gamma^{j}\right] / 2 $ and $\epsilon_{T}^{i j}=\epsilon^{-+i j}$. 
Furthermore, one can easily find that $E_{2}$, $G_{2}$, $F_{3}$ and $H_{3}$ are even functions of $\xi$, while $F_{2}$ and $H_{2}$ are odd functions of $\xi$. 
This implies $F_{2}(x,0,t)=0$, $H_{2}(x,0,t)=0$.
    
In the overlap representation, the quark-quark correlation function at $\xi=0$ can be expressed as~\cite{Sharma:2023ibp}
\begin{align}
    	\notag&F^{[\Gamma]}\left(x,0, t\right)  =\int \frac{d^{2} \boldsymbol{k}_{T}}{16 \pi^{3}} \sum_{\lambda_{q_{i}}} \sum_{\lambda_{q_{f}}} \\\notag&\times\left[\psi^{*}_{0}\left(x, \boldsymbol{k}_{T}^{\prime\prime},\lambda_{q_{f}},\uparrow\right) \psi_{0}\left(x, \boldsymbol{k}_{T}^{\prime},\lambda_{q_{i}},\uparrow\right)
    	\right.\\\notag&\left.+\psi^{*}_{0}\left(x, \boldsymbol{k}_{T}^{\prime\prime},\lambda_{q_{f}},\downarrow\right) \psi_{0}\left(x, \boldsymbol{k}_{T}^{\prime},\lambda_{q_{i}},\downarrow\right) \right]
    	\\&\times\frac{u_{\lambda_{q_{f}}}^{\dagger}\left(x P^{+}, \boldsymbol{k}_{T}-\frac{\boldsymbol{\Delta}_{T}}{2}\right) \gamma^{0} \Gamma u_{\lambda_{q_{i}}}\left(x P^{+}, \boldsymbol{k}_{T}+\frac{\boldsymbol{\Delta}_{T}}{2}\right)}{2 x P^{+}},
\end{align}
where $\lambda_{q_{i}}$ and $\lambda_{q_{f}}$ represent the initial and final state of the quark helicity, respectively. 
The spinor product $u_{\lambda_{q_{f}}}^{\dagger}\left(x P^{+}, \boldsymbol{k}_{T}-\frac{\boldsymbol{\Delta}_{T}}{2}\right) \gamma^{0} \Gamma u_{\lambda_{q_{i}}}\left(x P^{+}, \boldsymbol{k}_{T}+\frac{\boldsymbol{\Delta}_{T}}{2}\right)$ corresponds to the higher-twist Dirac matrices and encodes struck quark helicity combinations, 
and
\begin{align}	\nonumber\boldsymbol{k}^{{\prime}{\prime}}_T&
=\boldsymbol{k}_T-\frac{1}{2}(1-x)\boldsymbol{\Delta}_T  \displaybreak[0] \\ \boldsymbol{k}^{\prime}_T&=\boldsymbol{k}_T+\frac{1}{2}(1-x)\boldsymbol{\Delta}_T,\displaybreak[0]
    \end{align}	
are the transverse momenta for the final and initial struck quark $q$.
    
In this way, the overlap representation of the twist-3 GPDs for the pion and kaon in terms of the LCWFs can be expressed as
    \begin{widetext}
    \begin{align}
    	\notag
    	E_{2}^{\pi(K)}(x, 0, t) & = \int \frac{d^{2} \boldsymbol{k}_{T}}{16 \pi^{3}}\left[\frac{m_{q}}{xM_{\pi(K)}}\left(\psi^{\pi(K) *}_{0}\left(x, \boldsymbol{k}_{T}^{\prime\prime}, \uparrow, \uparrow\right) \psi^{\pi(K) }_{0}\left(x, \boldsymbol{k}_{T}^{\prime}, \uparrow, \uparrow\right)+\psi^{\pi(K) *}_{0}\left(x, \boldsymbol{k}_{T}^{\prime\prime}, \uparrow, \downarrow\right) \psi^{\pi(K) }_{0}\left(x, \boldsymbol{k}_{T}^{\prime}, \uparrow, \downarrow\right)\right.\right.   \displaybreak[0] \\\notag
    	& \left.+\psi^{\pi(K) *}_{0}\left(x, \boldsymbol{k}_{T}^{\prime\prime}, \downarrow, \uparrow\right) \psi^{\pi(K) }_{0}\left(x, \boldsymbol{k}_{T}^{\prime}, \downarrow, \uparrow\right)+\psi^{\pi(K) *}_{0}\left(x, \boldsymbol{k}_{T}^{\prime\prime}, \downarrow, \downarrow\right) \psi^{\pi(K) }_{0}\left(x, \boldsymbol{k}_{T}^{\prime}, \downarrow, \downarrow\right)\right) \displaybreak[0]\\ \notag
    	&+\frac{-\Delta_{1}+i\Delta_{2}}{2xM_{\pi(K)}}\left(\psi^{\pi(K) *}_{0}\left(x, \boldsymbol{k}_{T}^{\prime\prime}, \uparrow, \uparrow\right) \psi^{\pi(K) }_{0}\left(x, \boldsymbol{k}_{T}^{\prime}, \downarrow, \uparrow\right)+\psi^{\pi(K) *}_{0}\left(x, \boldsymbol{k}_{T}^{\prime\prime}, \uparrow, \downarrow\right) \psi^{\pi(K) }_{0}\left(x, \boldsymbol{k}_{T}^{\prime}, \downarrow, \downarrow\right)\right) \displaybreak[0]\\
    	& \left.+\frac{\Delta_{1}+i\Delta_{2}}{2xM_{\pi(K)}}\left(\psi^{\pi(K) *}_{0}\left(x, \boldsymbol{k}_{T}^{\prime\prime}, \downarrow ,\uparrow\right) \psi^{\pi(K) }_{0}\left(x, \boldsymbol{k}_{T}^{\prime}, \uparrow, \uparrow\right)+\psi^{\pi(K) *}_{0}\left(x, \boldsymbol{k}_{T}^{\prime\prime}, \downarrow, \downarrow\right) \psi^{\pi(K) }_{0}\left(x, \boldsymbol{k}_{T}^{\prime}, \uparrow, \downarrow\right)\right)\right], \displaybreak[0]\label{E2}
    \\
    	\label{G2}\notag 
    	\frac{-i\Delta_{2}}{M_{\pi(K)}}G_{2}^{\pi(K)}(x, 0, t) & = \int \frac{d^{2} \boldsymbol{k}_{T}}{16 \pi^{3}}\left[\frac{m_{q}}{2xM_{\pi(K)}}\left(\psi^{\pi(K) *}_{0}\left(x, \boldsymbol{k}_{T}^{\prime\prime}, \uparrow, \uparrow\right) \psi^{\pi(K) }_{0}\left(x, \boldsymbol{k}_{T}^{\prime}, \downarrow, \uparrow\right)+\psi^{\pi(K) *}_{0}\left(x, \boldsymbol{k}_{T}^{\prime\prime}, \uparrow, \downarrow\right) \psi^{\pi(K) }_{0}\left(x, \boldsymbol{k}_{T}^{\prime}, \downarrow, \downarrow\right)\right.\right. \displaybreak[0] \\ \notag
    	& \left.+\psi^{\pi(K) *}_{0}\left(x, \boldsymbol{k}_{T}^{\prime\prime}, \downarrow \uparrow,\right) \psi^{\pi(K) }_{0}\left(x, \boldsymbol{k}_{T}^{\prime}, \uparrow, \uparrow\right)+\psi^{\pi(K) *}_{0}\left(x, \boldsymbol{k}_{T}^{\prime\prime}, \downarrow, \downarrow\right) \psi^{\pi(K) }_{0}\left(x, \boldsymbol{k}_{T}^{\prime}, \uparrow, \downarrow\right)\right)\displaybreak[0] \\ \notag
    	&+\frac{2k_{1}+i\Delta_{2}}{2xM_{\pi(K)}}\left(\psi^{\pi(K) *}_{0}\left(x, \boldsymbol{k}_{T}^{\prime\prime}, \uparrow, \uparrow\right) \psi^{\pi(K) }_{0}\left(x, \boldsymbol{k}_{T}^{\prime}, \uparrow, \uparrow\right)+\psi^{\pi(K) *}_{0}\left(x, \boldsymbol{k}_{T}^{\prime\prime}, \uparrow, \downarrow\right) \psi^{\pi(K) }_{0}\left(x, \boldsymbol{k}_{T}^{\prime}, \uparrow, \downarrow\right)\right) \displaybreak[0] \\
    	& \left.+\frac{-2k_{1}+i\Delta_{2}}{2xM_{\pi(K)}}\left(\psi^{\pi(K) *}_{0}\left(x, \boldsymbol{k}_{T}^{\prime\prime}, \downarrow ,\uparrow\right) \psi^{\pi(K) }_{0}\left(x, \boldsymbol{k}_{T}^{\prime}, \downarrow, \uparrow\right)+\psi^{\pi(K) *}_{0}\left(x, \boldsymbol{k}_{T}^{\prime\prime}, \downarrow, \downarrow\right) \psi^{\pi(K) }_{0}\left(x, \boldsymbol{k}_{T}^{\prime}, \downarrow, \downarrow\right)\right)\right]. \displaybreak[0]
    \end{align}
In the case of twist-4, the GPDs in the overlap representation can be expressed as
    \begin{align}
    	\notag
    	&F_{3}^{\pi(K)}(x, 0, t)
    	\displaybreak[0] \\ \notag&=  \int \frac{d^{2} \boldsymbol{k}_{T}}{16 \pi^{3}}\left[\frac{m_{q}^{2}+\boldsymbol{k}_{T}^{2}
    -\frac{1}{4}\boldsymbol{\Delta}_{T}^{2}+i\boldsymbol{k}_{T}
    \times\boldsymbol{\Delta}_{T}}{2x^{2}M^{2}_{\pi(K)}}\left(\psi^{\pi(K) *}_{0}\left(x, \boldsymbol{k}_{T}^{\prime\prime}, \uparrow, \uparrow\right) \psi^{\pi(K) }_{0}\left(x, \boldsymbol{k}_{T}^{\prime}, \uparrow, \uparrow\right)+\psi^{\pi(K) *}_{0}\left(x, \boldsymbol{k}_{T}^{\prime\prime}, \uparrow, \downarrow\right) \psi^{\pi(K) }_{0}\left(x, \boldsymbol{k}_{T}^{\prime}, \uparrow, \downarrow\right)\right)\right. \displaybreak[0]\\ \notag
    	& +\frac{m_{q}^{2}+\boldsymbol{k}_{T}^{2}-\frac{1}{4}\boldsymbol{\Delta}_{T}^{2}
    -i\boldsymbol{k}_{T}\times\boldsymbol{\Delta}_{T}}{2x^{2}M^{2}_{\pi(K)}}
    \left(\psi^{\pi(K) *}_{0}\left(x, \boldsymbol{k}_{T}^{\prime\prime}, \downarrow ,\uparrow\right) \psi^{\pi(K) }_{0}\left(x, \boldsymbol{k}_{T}^{\prime}, \downarrow, \uparrow\right)+\psi^{\pi(K) *}_{0}\left(x, \boldsymbol{k}_{T}^{\prime\prime}, \downarrow, \downarrow\right) \psi^{\pi(K) }_{0}\left(x, \boldsymbol{k}_{T}^{\prime}, \downarrow, \downarrow\right)\right) \displaybreak[0]\\ \notag    	&+\frac{m_q(-\Delta_{1}+i\Delta_{2})}{2x^{2}M^{2}_{\pi(K)}}\left(\psi^{\pi(K) *}_{0}\left(x, \boldsymbol{k}_{T}^{\prime\prime}, \uparrow, \uparrow\right) \psi^{\pi(K) }_{0}\left(x, \boldsymbol{k}_{T}^{\prime}, \downarrow, \uparrow\right)+\psi^{\pi(K) *}_{0}\left(x, \boldsymbol{k}_{T}^{\prime\prime}, \uparrow, \downarrow\right) \psi^{\pi(K) }_{0}\left(x, \boldsymbol{k}_{T}^{\prime}, \downarrow, \downarrow\right)\right) \displaybreak[0] \\
    	& \left.+\frac{m_q(\Delta_{1}+i\Delta_{2})}{2x^{2}M^{2}_{\pi(K)}}\left(\psi^{\pi(K) *}_{0}\left(x, \boldsymbol{k}_{T}^{\prime\prime}, \downarrow, \uparrow\right) \psi^{\pi(K) }_{0}\left(x, \boldsymbol{k}_{T}^{\prime}, \uparrow, \uparrow\right)+\psi^{\pi(K) *}_{0}\left(x, \boldsymbol{k}_{T}^{\prime\prime}, \downarrow, \downarrow\right) \psi^{\pi(K) }_{0}\left(x, \boldsymbol{k}_{T}^{\prime}, \uparrow, \downarrow\right)\right)\right], \displaybreak[0]\label{F3}\\
    	\notag\label{H3}
    	&\frac{-i\Delta_{2}}{M_{\pi(K)}}H_{3}^{\pi(K)}(x, 0, t)\displaybreak[0] \\ 
    \notag&=  \int \frac{d^{2} \boldsymbol{k}_{T}}{16 \pi^{3}}\left[\frac{m_{q}(2k_{1}+i\Delta_{2})}{2x^{2}M^{2}_{\pi(K)}}\left(\psi^{\pi(K) *}_{0}\left(x, \boldsymbol{k}_{T}^{\prime\prime}, \uparrow, \uparrow\right) \psi^{\pi(K) }_{0}\left(x, \boldsymbol{k}_{T}^{\prime}, \uparrow, \uparrow\right)+\psi^{\pi(K) *}_{0}\left(x, \boldsymbol{k}_{T}^{\prime\prime}, \uparrow, \downarrow\right) \psi^{\pi(K) }_{0}\left(x, \boldsymbol{k}_{T}^{\prime}, \uparrow, \downarrow\right)\right)\right. \displaybreak[0]\\ \notag
    	& +\frac{m_{q}(-2k_{1}+i\Delta_{2})}{2x^{2}M^{2}_{\pi(K)}}\left(\psi^{\pi(K) *}_{0}\left(x, \boldsymbol{k}_{T}^{\prime\prime}, \downarrow ,\uparrow\right) \psi^{\pi(K) }_{0}\left(x, \boldsymbol{k}_{T}^{\prime}, \downarrow, \uparrow\right)+\psi^{\pi(K) *}_{0}\left(x, \boldsymbol{k}_{T}^{\prime\prime}, \downarrow, \downarrow\right) \psi^{\pi(K) }_{0}\left(x, \boldsymbol{k}_{T}^{\prime}, \downarrow, \downarrow\right)\right) \displaybreak[0] \\ \notag    	&+\frac{m_{q}^{2}-(k_{1}-ik_{2})^{2}+\frac{1}{4}(\Delta_{1}-i\Delta_{2})^2}
    {2x^{2}M^{2}_{\pi(K)}}\left(\psi^{\pi(K) *}_{0}\left(x, \boldsymbol{k}_{T}^{\prime\prime}, \uparrow, \uparrow\right) \psi^{\pi(K) }_{0}\left(x, \boldsymbol{k}_{T}^{\prime}, \downarrow, \uparrow\right)+\psi^{\pi(K) *}_{0}\left(x, \boldsymbol{k}_{T}^{\prime\prime}, \uparrow, \downarrow\right) \psi^{\pi(K) }_{0}\left(x, \boldsymbol{k}_{T}^{\prime}, \downarrow, \downarrow\right)\right) \displaybreak[0]\\
    	& \left.+\frac{m_{q}^{2}-(k_{1}+ik_{2})^{2}+\frac{1}{4}(\Delta_{1}+i\Delta_{2})^2}
    {2x^{2}M^{2}_{\pi(K)}}\left(\psi^{\pi(K) *}_{0}\left(x, \boldsymbol{k}_{T}^{\prime\prime}, \downarrow, \uparrow\right) \psi^{\pi(K) }_{0}\left(x, \boldsymbol{k}_{T}^{\prime}, \uparrow, \uparrow\right)+\psi^{\pi(K) *}_{0}\left(x, \boldsymbol{k}_{T}^{\prime\prime}, \downarrow, \downarrow\right) \psi^{\pi(K) }_{0}\left(x, \boldsymbol{k}_{T}^{\prime}, \uparrow, \downarrow\right)\right)\right]. \displaybreak[0]
\end{align}

Using the LCWFs in Eq.~(\ref{piLCWFs}) and Eq.~(\ref{KLCWFs}) as well as the overlap representation in Eqs.~(\ref{E2}-\ref{H3}), we obtain the analytic results for the twist-3 and twist-4 GPDs of the pion and kaon:
\begin{align}
E_{2}^{\pi}(x, 0, t)&=\frac{m}{xM_{\pi}}\int \frac{d^2 \boldsymbol{k}_{T}}{16 \pi^{3}}\left[m^{2}+ \boldsymbol{k}_{T}^{2}-\frac{1}{4}(1-x)^{2}\boldsymbol{\Delta}^{2}_{T}+2m(1-x)
\boldsymbol{\Delta}^{2}_{T}\right] \frac{\varphi^{\pi*}\left(x, \boldsymbol{k}_{T}^{\prime\prime}\right)\varphi^{\pi}\left(x, \boldsymbol{k}_{T}^{\prime}\right)}{\omega_{+}^{\pi} \omega_{-}^{\pi}},\displaybreak[0]\\
G_{2}^{\pi}(x, 0, t)&=-\frac{1}{2x}\int \frac{d^2 \boldsymbol{k}_{T}}{16 \pi^{3}}\left[m^{2}+ \boldsymbol{k}_{T}^{2}-\frac{1}{4}(1-x)^{2}\boldsymbol{\Delta}^{2}_{T}-m^{2}(1-x)
-\frac{2k_{1}}{\Delta_{2}}(1-x)\boldsymbol{k}_{T}\times\boldsymbol{\Delta}_{T}\right] \frac{\varphi^{\pi*}\left(x, \boldsymbol{k}_{T}^{\prime\prime}\right)\varphi^{\pi}\left(x, \boldsymbol{k}_{T}^{\prime}\right)}{\omega_{+}^{\pi} \omega_{-}^{\pi}}, \displaybreak[0]\\ \notag F_{3}^{\pi}(x, 0, t)&=\frac{1}{2x^{2}M_{\pi}^{2}}\int \frac{d^2 \boldsymbol{k}_{T}}{16 \pi^{3}}\left[(m^{2}+ \boldsymbol{k}_{T}^{2}-\frac{1}{4}\boldsymbol{\Delta}^{2}_{T})\left(m^{2}+ \boldsymbol{k}_{T}^{2}-\frac{1}{4}(1-x)^{2}\boldsymbol{\Delta}^{2}_{T}\right)
+(1-x)(\boldsymbol{k}_{T}\times\boldsymbol{\Delta}_{T})^{2} \right.\displaybreak[0] \\
&\left.+m^{2}(1-x)\boldsymbol{\Delta}_{T}^{2}\right]
\frac{\varphi^{\pi*}\left(x, \boldsymbol{k}_{T}^{\prime\prime}\right)\varphi^{\pi}\left(x, \boldsymbol{k}_{T}^{\prime}\right)}{\omega_{+}^{\pi} \omega_{-}^{\pi}}, \displaybreak[0]\\
H_{3}^{\pi}(x, 0, t)&=\frac{m}{2x^{2}M_{\pi}}\int \frac{d^{3}\boldsymbol{k}_{T}^{2}}{16 \pi^{3}}\left[(1-x)(m^{2}+\boldsymbol{k}_{T}^{2}-\frac{1}{4}\boldsymbol{\Delta}^{2}_{T})
-\left(m^{2}+\boldsymbol{k}_{T}^{2}
-\frac{1}{4}(1-x)^{2}\boldsymbol{\Delta}^{2}_{T}\right)\right]\frac{\varphi^{\pi*}
\left(x,\boldsymbol{k}_{T}^{\prime\prime}\right)\varphi^{\pi}
\left(x,\boldsymbol{k}_{T}^{\prime}\right)}{\omega_{+}^{\pi} \omega_{-}^{\pi}},\displaybreak[0]\\
\notag E_{2}^{K}(x, 0, t)&=\frac{1}{2xM_{K}}\int \frac{d^2 \boldsymbol{k}_{T}}{16 \pi^{3}}\left[2m_{1}\left(\left((1-x)m_{1}+xm_{2}\right)^{2}+ \boldsymbol{k}_{T}^{2}-\frac{1}{4}(1-x)^{2}\boldsymbol{\Delta}^{2}_{T}\right)
\right.\displaybreak[0]\\
&\left.+(1-x)\left((1-x)m_{1}+xm_{2}\right)\boldsymbol{\Delta}^{2}_{T}\right]
\frac{\varphi^{K*}\left(x,\boldsymbol{k}_{T}^{\prime\prime}\right)\varphi^{K}\left(x, \boldsymbol{k}_{T}^{\prime}\right)}{\omega_{+}^{K} \omega_{-}^{K}},\displaybreak[0]\\ \notag G_{2}^{K}(x, 0, t)&=-\frac{1}{2x}\int \frac{d^2 \boldsymbol{k}_{T}}{16 \pi^{3}}\left[\left((1-x)m_{1}+xm_{2}\right)^{2}+ \boldsymbol{k}_{T}^{2}-\frac{1}{4}(1-x)^{2}\boldsymbol{\Delta}^{2}_{T}
-m_{1}(1-x)\left((1-x)m_{1}+xm_{2}\right) 	 \notag\right. \displaybreak[0]\\
&\left.-\frac{2k_{1}}{\Delta_{2}}(1-x)\boldsymbol{k}_{T}\times\boldsymbol{\Delta}_{T}\right] \frac{\varphi^{K*}\left(x, \boldsymbol{k}_{T}^{\prime\prime}\right)\varphi^{K}\left(x, \boldsymbol{k}_{T}^{\prime}\right)}{\omega_{+}^{K} \omega_{-}^{K}},\displaybreak[0]\\ \notag F_{3}^{K}(x, 0, t)&=\frac{1}{2x^{2}M_{K}^{2}}\int \frac{d^2 \boldsymbol{k}_{T}}{16 \pi^{3}}\left[(m_{1}^{2}+ \boldsymbol{k}_{T}^{2} -\frac{1}{4}\boldsymbol{\Delta}^{2}_{T})\left(((1-x)m_{1}+xm_{2})^{2}+\boldsymbol{k}_{T}^{2}
-\frac{1}{4}(1-x)^{2}\boldsymbol{\Delta}^{2}_{T}\right)\right.\displaybreak[0]\\
&\left.+(1-x)(\boldsymbol{k}_{T}\times\boldsymbol{\Delta}_{T})^{2}
+m_{1}((1-x)m_{1}+xm_{2})(1-x)\boldsymbol{\Delta}_{T}^{2}\right]\frac{\varphi^{K*}\left(x, \boldsymbol{k}_{T}^{\prime\prime}\right)\varphi^{K}\left(x, \boldsymbol{k}_{T}^{\prime}\right)}{\omega_{+}^{K} \omega_{-}^{K}},\displaybreak[0]\\ \notag H_{3}^{K}(x, 0, t)&=\frac{1}{2x^{2}M_{K}}\int \frac{d^2 \boldsymbol{k}_{T}}{16 \pi^{3}}\left[(1-x)((1-x)m_{1}+xm_{2})(m_{1}^{2}+\boldsymbol{k}_{T}^{2}
-\frac{1}{4}\boldsymbol{\Delta}^{2}_{T})\right.\displaybreak[0]\\
&\left.-m_{1}\left(((1-x)m_{1}+xm_{2})^{2}+\boldsymbol{k}_{T}^{2}
-\frac{1}{4}(1-x)^{2}\boldsymbol{\Delta}^{2}_{T}\right)-\frac{2k_{1}}{\Delta_{2}}
x(1-x)(m_{1}+m_{2})\right]\frac{\varphi^{K*}\left(x,\boldsymbol{k}_{T}^{\prime\prime}\right)
\varphi^{K}\left(x, \boldsymbol{k}_{T}^{\prime}\right)}{\omega_{+}^{K} \omega_{-}^{K}},\displaybreak[0]
\end{align}
\end{widetext}
where
   \begin{align}
 	\omega_{+}^{\pi} & = \left[\left(\boldsymbol{k}_{T}+(1-x) \frac{\boldsymbol{\Delta}_{T}}{2}\right)^{2}+m^{2}\right]^{\frac{1}{2}}, \\
 	\omega_{-}^{\pi} & = \left[\left(\boldsymbol{k}_{T}-(1-x) \frac{\boldsymbol{\Delta}_{T}}{2}\right)^{2}+m^{2}\right]^{\frac{1}{2}},\\
 	\omega_{+}^{K} & = \left[\left(\boldsymbol{k}_{T}+(1-x) \frac{\boldsymbol{\Delta}_{T}}{2}\right)^{2}+L^{2}\right]^{\frac{1}{2}}, \\
 	\omega_{-}^{K} & = \left[\left(\boldsymbol{k}_{T}-(1-x) \frac{\boldsymbol{\Delta}_{T}}{2}\right)^{2}+L^{2}\right]^{\frac{1}{2}}.
   \end{align}

\section{Numerical results and discussion}\label{Sec4}

\subsection{Numerical results for twist-3 and twist-4 GPDs of pion and kaon}

\begin{figure*}[htbp]
	\centering
	\subfigure{\begin{minipage}[b]{0.42\linewidth}
			\centering
			\includegraphics[width=\linewidth]{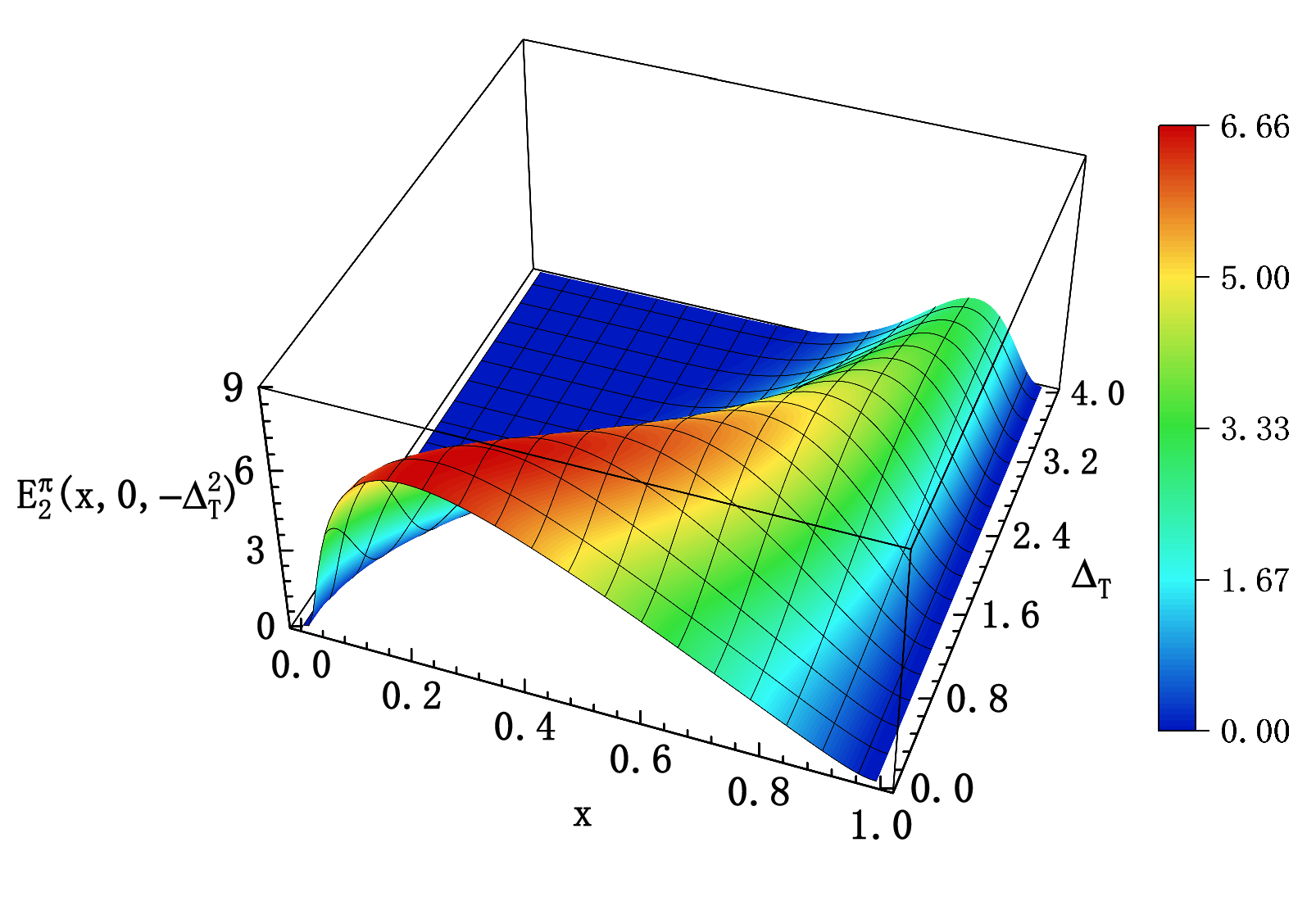}
	\end{minipage}}
	\subfigure{\begin{minipage}[b]{0.42\linewidth}
			\centering
			\includegraphics[width=\linewidth]{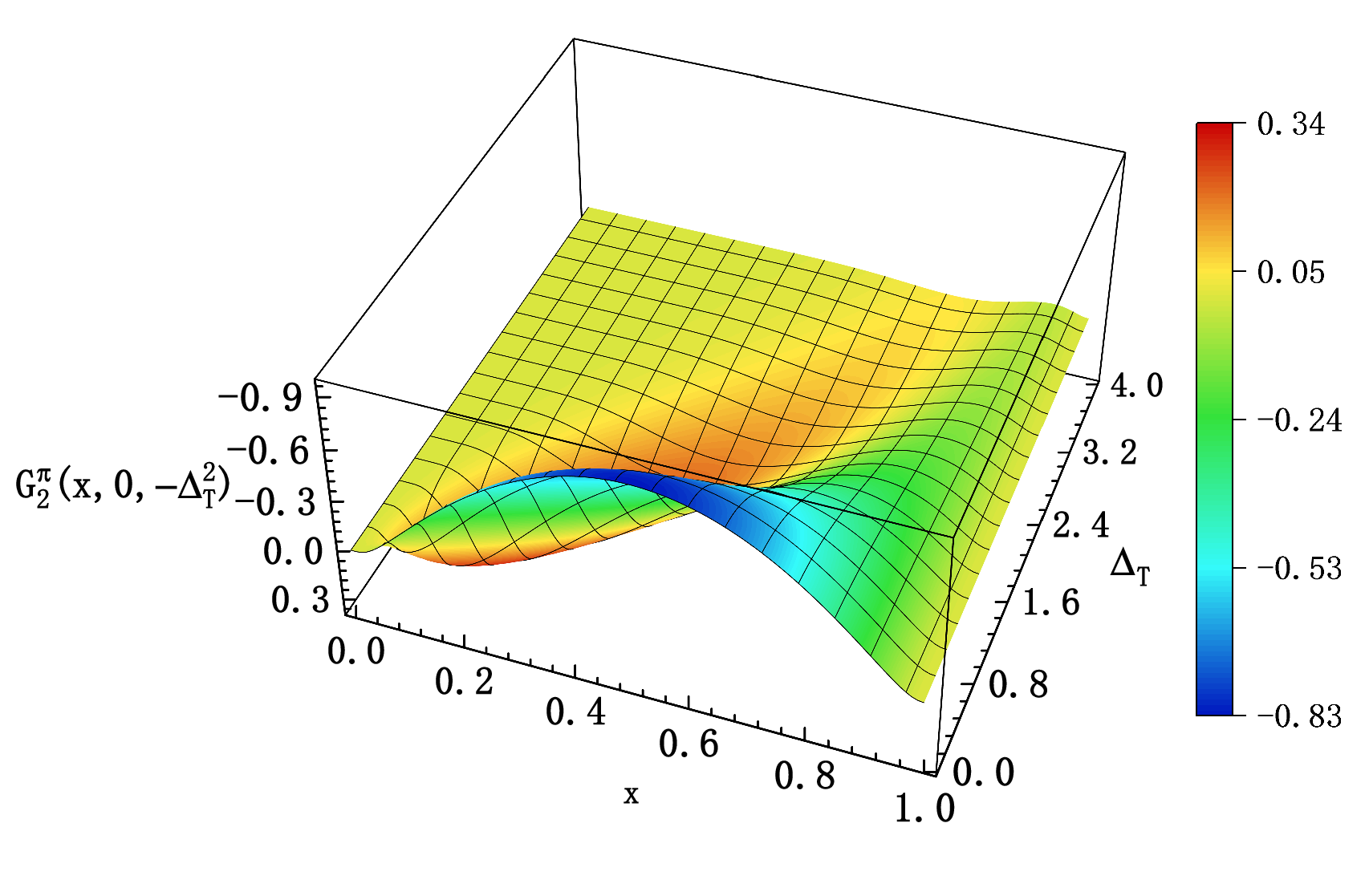}    	
	\end{minipage}}
	\subfigure{\begin{minipage}[b]{0.42\linewidth}
			\centering
			\includegraphics[width=\linewidth]{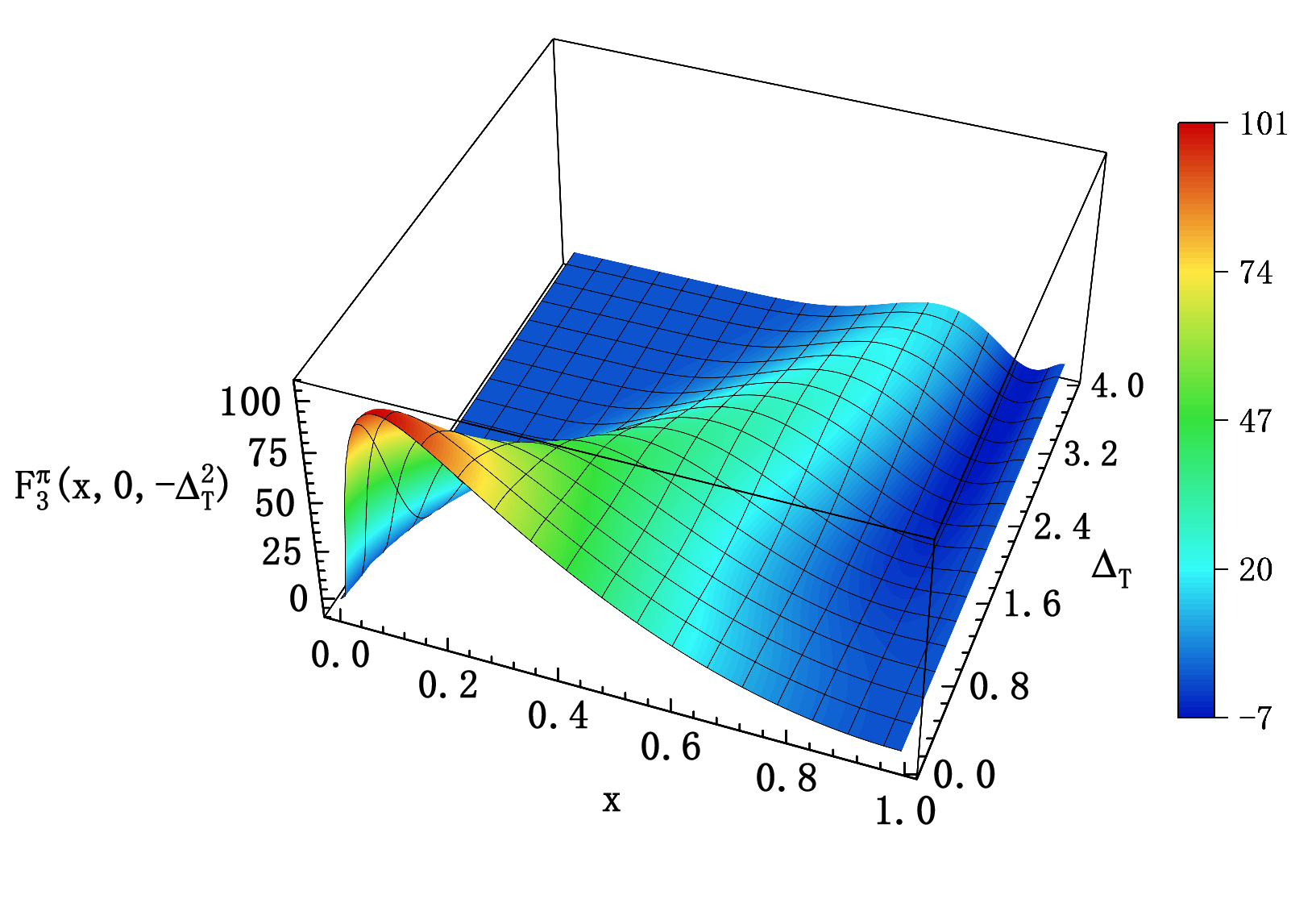}
	\end{minipage}}
	\subfigure{\begin{minipage}[b]{0.42\linewidth}
			\centering
			\includegraphics[width=\linewidth]{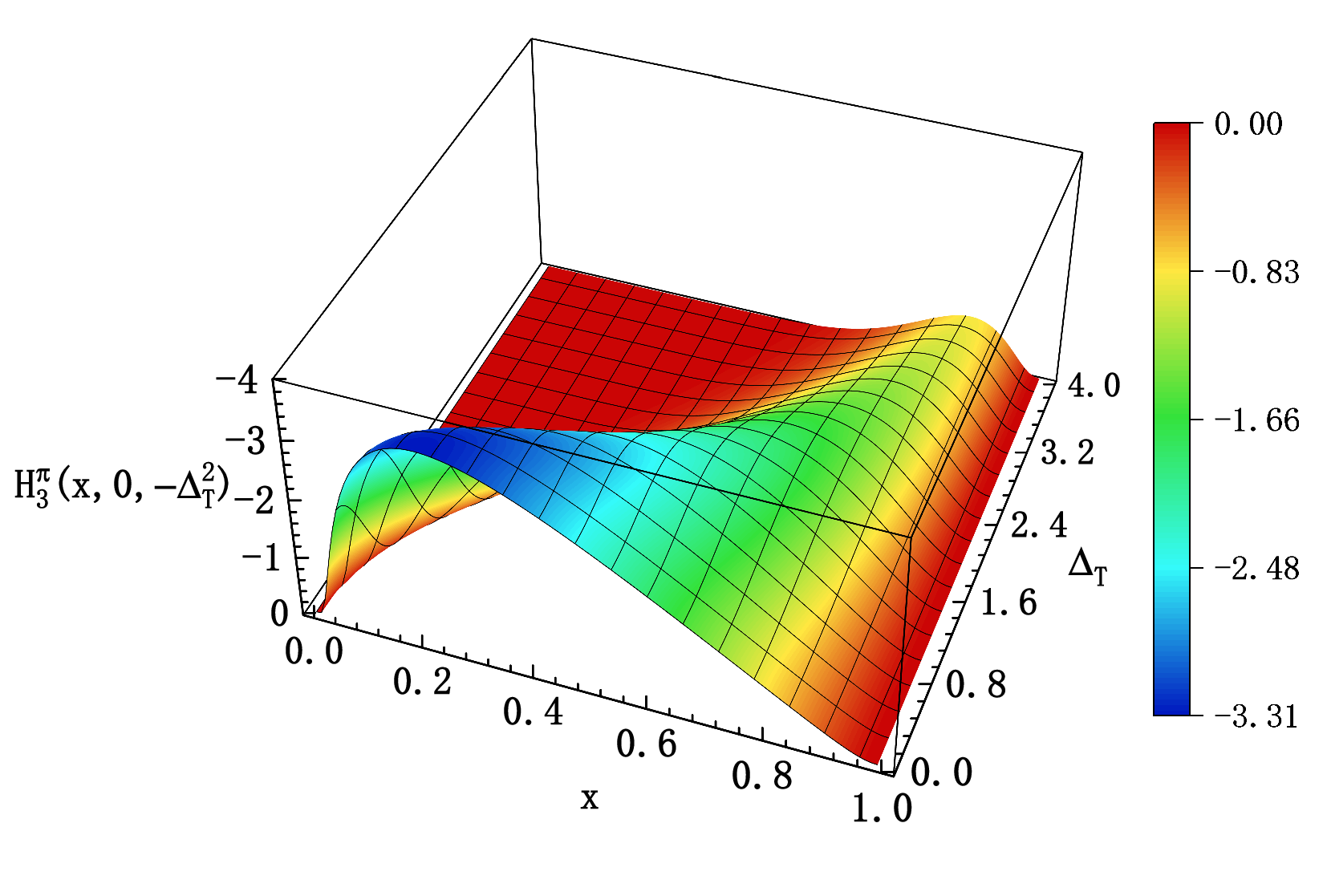}    	
	\end{minipage}}
	\caption{The twist-3 (upper panel) and twist-4 (lower panel) GPDs for the pion meson with respect to $x$ and ${\Delta}_{T}$ in the LCQM.} 
	\label{piGPDs}      
\end{figure*}

\begin{figure*}[htbp]
	\centering
	\subfigure{\begin{minipage}[b]{0.42\linewidth}
			\centering
			\includegraphics[width=\linewidth]{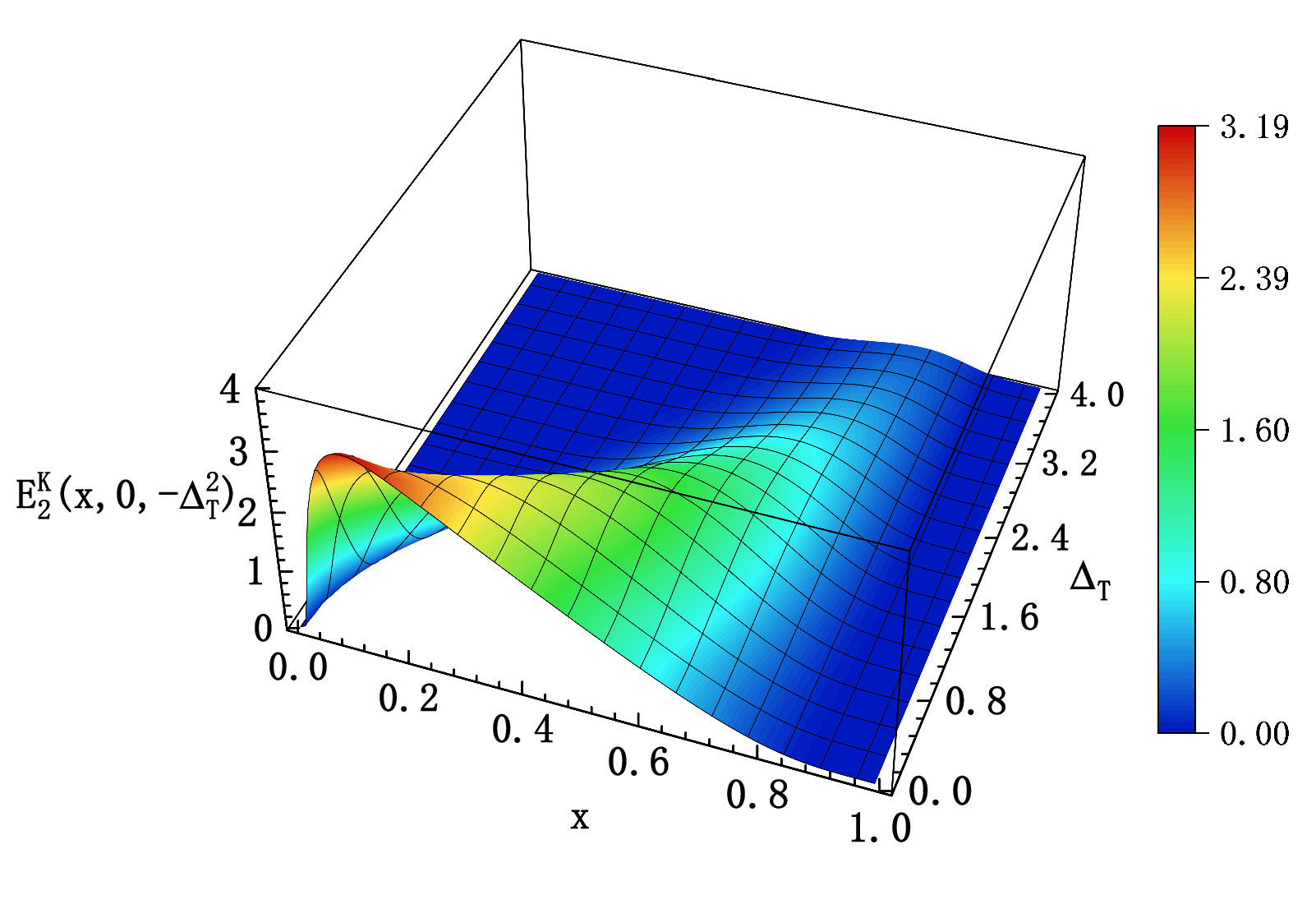}
	\end{minipage}}
	\subfigure{\begin{minipage}[b]{0.42\linewidth}
			\centering
			\includegraphics[width=\linewidth]{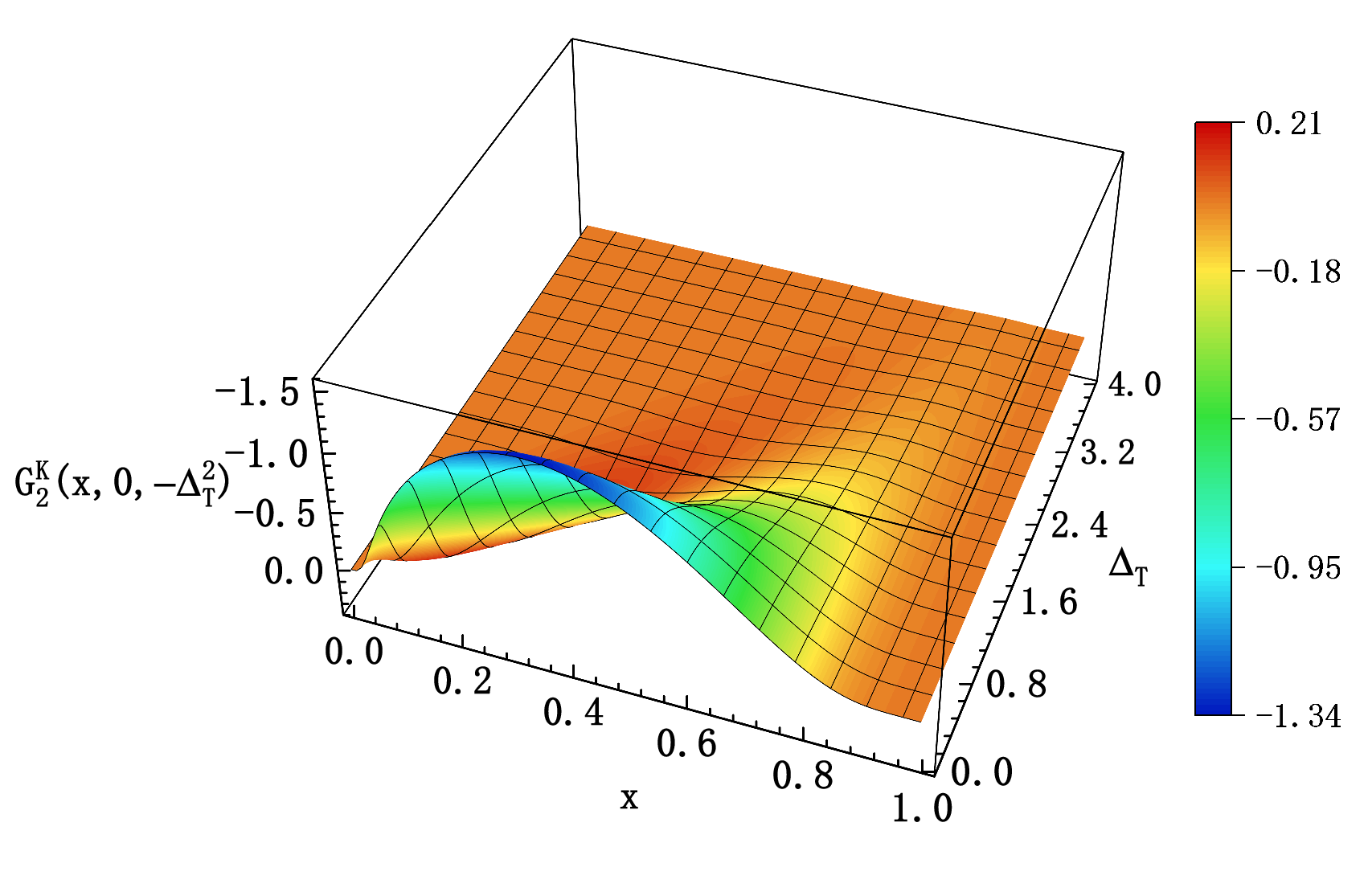}    	
	\end{minipage}}
	\subfigure{\begin{minipage}[b]{0.42\linewidth}
			\centering
			\includegraphics[width=\linewidth]{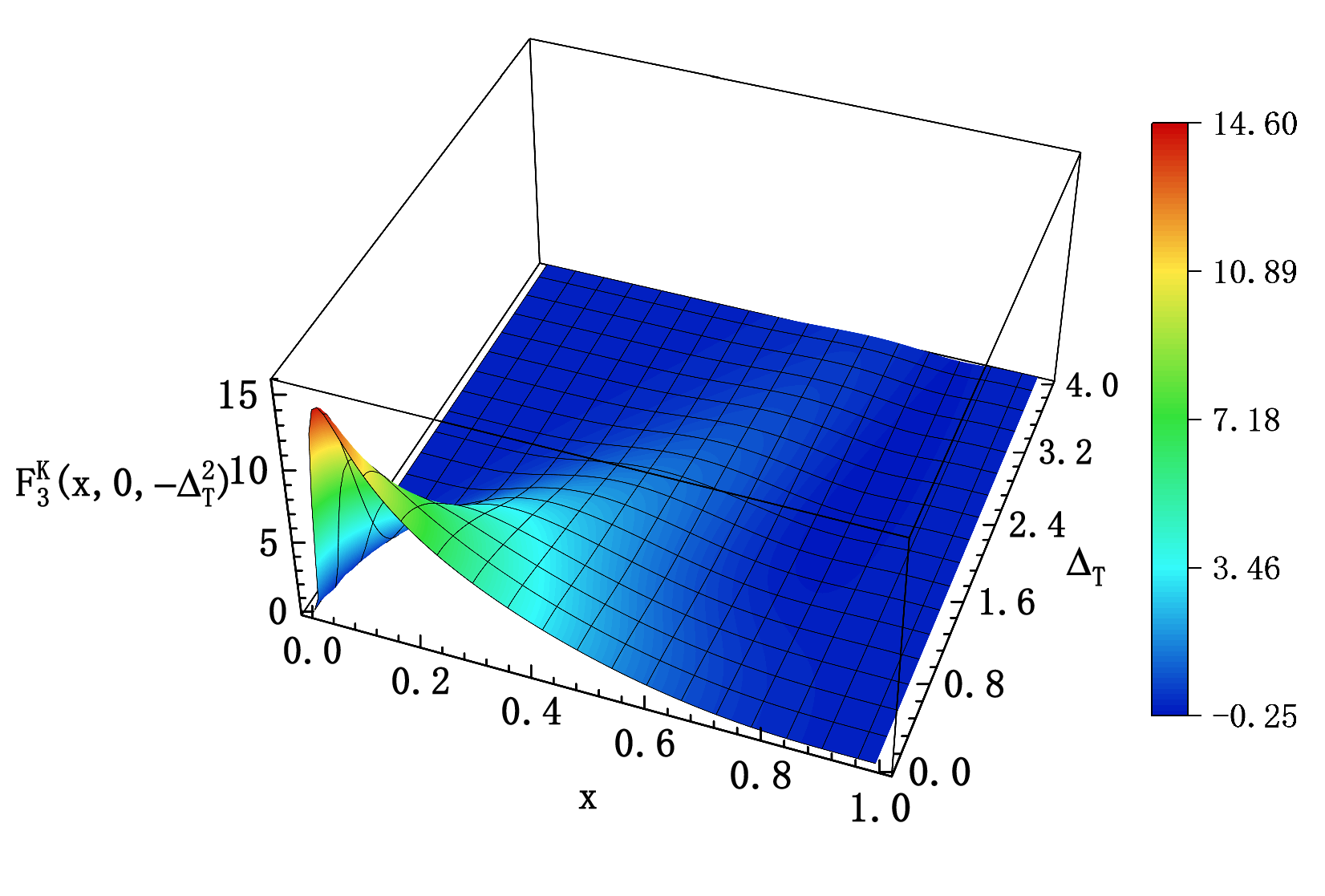}
	\end{minipage}}
	\subfigure{\begin{minipage}[b]{0.42\linewidth}
			\centering
			\includegraphics[width=\linewidth]{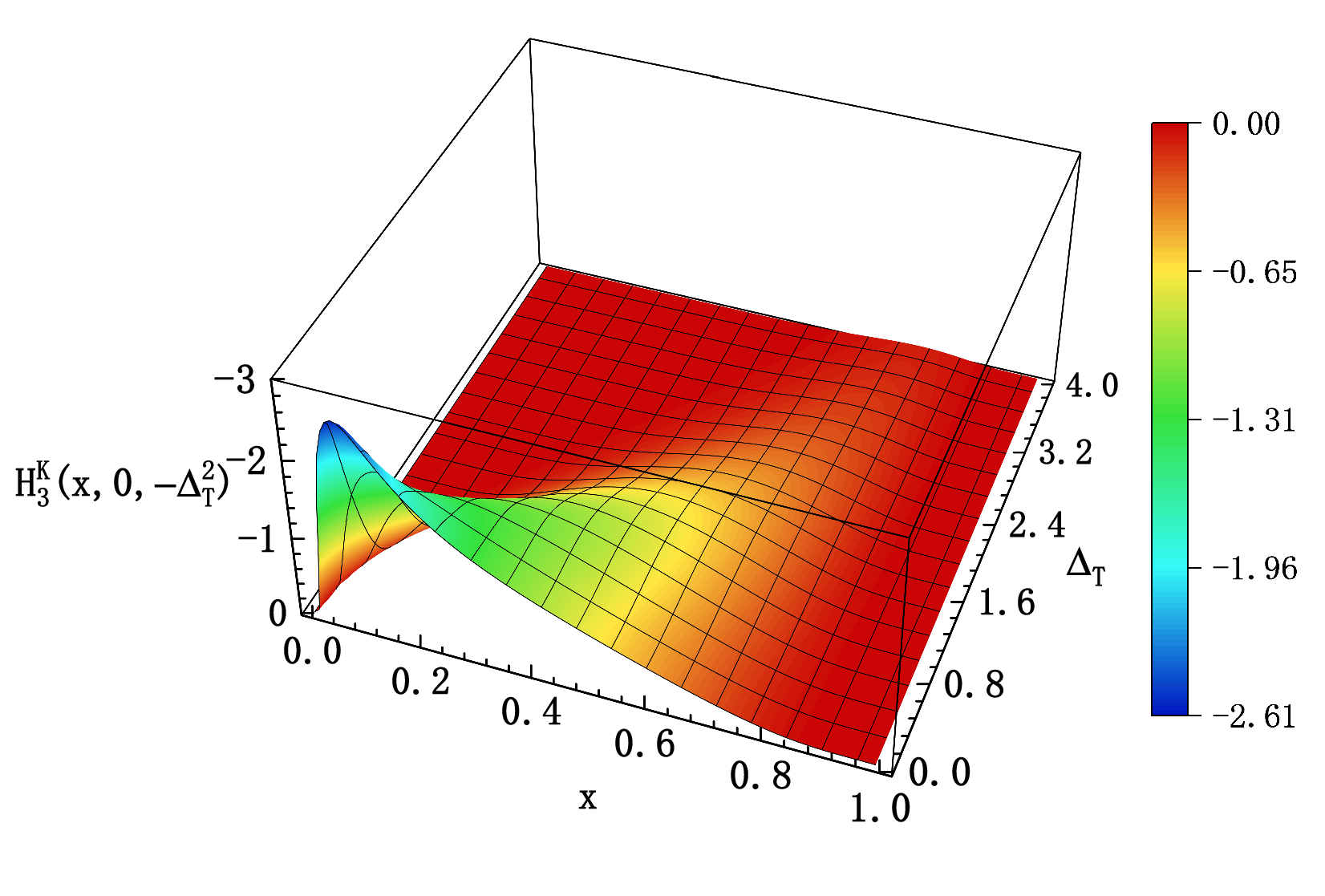}    	
	\end{minipage}}
	\caption{The twist-3 (upper panel) and twist-4 (lower panel) GPDs of the $u$ quark for $K^+$ with respect to $x$ and ${\Delta}_{T}$ in the LCQM.} 
	\label{KGPDs}      
\end{figure*}

In this section, we present the numerical results for the twist-3 and twist-4 GPDs for the pion and kaon. 
To specify the values of the parameters, namely, the constituent quark masses and the harmonic scale $\beta$ in our model, we adopt the values shown in Table~\ref{tab1} from Ref.~\cite{Qian:2008px}, where the parameters are determined by the experimental data.
Using these parameters, the normalization constants are found to be $A^\pi=44.27$, $A^K=47.19$.

 \begin{center}\label{tab1}
	\setlength{\tabcolsep}{5mm}
	\renewcommand\arraystretch{1.5}
	\begin{tabular}{ c | c | c }
		\hline
		Meson & Mass(GeV) & $\beta$(GeV) \\
		\hline
		\hline
		$\pi^+(u\bar{d})$ & $m=0.2$ & 0.410 \\
		\hline
		$K^+(u\bar{s})$ & $m_{1}=0.2, m_{2}=0.556$ & 0.405 \\
		\hline
	\end{tabular}
	\captionof{table}{Values of the parameters obtained from Ref.~\cite{Qian:2008px}.} \label{tab1}
\end{center}

In order to show the $x$-dependence and ${\Delta}_{T}$-dependence of the twist-3 and twist-4 GPDs for the pion and kaon simultaneously, we present the 3-D shapes of these distributions in Fig.~\ref{piGPDs} and Fig.~\ref{KGPDs}. 
In the upper and lower panels of Fig.~\ref{piGPDs}, we plot the twist-3 GPDs $E^{\pi}_{2}(x,0,-\boldsymbol{\Delta}_{T}^{2})$, $G^{\pi}_{2}(x,0,-\boldsymbol{\Delta}_{T}^{2})$ and the twist-4 GPDs $F^{\pi}_{3}(x,0,-\boldsymbol{\Delta}_{T}^{2})$, $H^{\pi}_{3}(x,0,-\boldsymbol{\Delta}_{T}^{2})$ as functions of $x$ and $\Delta_T$, respectively. 
First of all, $E^{\pi}_{2}(x,0,-\boldsymbol{\Delta}_{T}^{2})$ is positive in the entire $x$ and  ${\Delta}_{T}$ region. 
We find the magnitude of $E^{\pi}_{2}(x,0,-\boldsymbol{\Delta}_{T}^{2})$ decreases with increasing ${\Delta}_{T}$. 
Moreover, with increases of $x$, $E^{\pi}_{2}(x,0,-\boldsymbol{\Delta}_{T}^{2})$  increases at first and then decreases. The maxima of $E^{\pi}_{2}(x,0,-\boldsymbol{\Delta}_{T}^{2})$ is driving to larger x region with increasing $\Delta_T$. 
Unlike $E^{\pi}_{2}(x,0,-\boldsymbol{\Delta}_{T}^{2})$, the twist-3 GPDs $G^{\pi}_{2}(x,0,-\boldsymbol{\Delta}_{T}^{2})$ possesses both positive and negative values with the variation of $x$ and ${\Delta}_{T}$.
Besides, $G^{\pi}_{2}(x,0,-\boldsymbol{\Delta}_{T}^{2})$ decreases rapidly with increasing ${\Delta}_T$. 
As $x$ increases, $G^{\pi}_{2}(x,0,-\boldsymbol{\Delta}_{T}^{2})$ turns to positive from negative value, indicating a sign flips in the distribution at a specific value of $x$ and  ${\Delta}_{T}$. 
For twist-4 GPDs, we find that both $F^{\pi}_{3}(x,0,-\boldsymbol{\Delta}_{T}^{2})$ and $H^{\pi}_{3}(x,0,-\boldsymbol{\Delta}_{T}^{2})$ have maximum value at  $\Delta_T=0$.  
As $x$ increases, the peaks of $F^{\pi}_{3}(x,0,-\boldsymbol{\Delta}_{T}^{2})$ and $H^{\pi}_{3}(x,0,-\boldsymbol{\Delta}_{T}^{2})$ shift from lower ${\Delta}_{T}$ to higher ${\Delta}_{T}$. 
The sign of $H^{\pi}_{3}(x,0,-\boldsymbol{\Delta}_{T}^{2})$ is negative in the entire region of $x$ and ${\Delta}_{T}$, while $F^{\pi}_{3}(x,0,-\boldsymbol{\Delta}_{T}^{2})$ is positive.

In Fig.~\ref{KGPDs}, we plot the twist-3 and twist-4 GPDs of $u$ quark for the kaon meson vs $x$ and ${\Delta}_{T}$. 
We can clearly see that $E^{K}_{2}(x,0,-\boldsymbol{\Delta}_{T}^{2})$ and $G^{K}_{2}(x,0,-\boldsymbol{\Delta}_{T}^{2})$ show similar behaviors to those of the pion. 
It is also found that, within any plot,  GPDs decreases with the value of ${\Delta}_{T}$ increases. The signs of the twist-3 GPDs $E_{2}(x,0,-\boldsymbol{\Delta}_{T}^{2})$, $G_{2}(x,0,-\boldsymbol{\Delta}_{T}^{2})$ and twist-4 GPDs $F_{3}(x,0,-\boldsymbol{\Delta}_{T}^{2})$, $H_{3}(x,0,-\boldsymbol{\Delta}_{T}^{2})$ are same for the both cases of pion and kaon. Compared to the pion, the distributions of twist-3 and twist-4 GPDs of kaon peak at smaller $x$ and their peak amplitudes are smaller.

As there are no other calculations on the twist-3 and twist-4 GPDs for the pion and kaon mesons so far, we compare our results to those for the proton for qualitative discussion.
In Ref.~\cite{Zhang:2023xfe}, the twist-3 GPDs $E_2^{q/p}(x,0,-\boldsymbol{\Delta}_{T}^{2})$ (denoted by $H_2(x,0,-t)$ in their paper ) was calculated in BLFQ.
We find that the size and sign of $E_2^{\pi}$ or $E_2^{K}$ in our model are similar to those of $E_2^{q/p}$ in Ref.~\cite{Zhang:2023xfe},  while the $x$-dependence and $\Delta_T$-dependence of the two GPDs are different. 
In our model  $E_2^{\pi}$  is zero at $x$ approaches zero, while $E_2^{q/p}$ in BLFQ has the largest size in the $x\rightarrow 0$ region.
In Ref.~\cite{Sharma:2023ibp} the twist-4 GPD $F_3^{q/p}(x,0,-\boldsymbol{\Delta}_{T}^{2})$ (denoted by $H_3(x,0,-t)$ in their paper ) was calculated within LFQDM.
It is found that the shape and the sign of $F_3^{u/p}$ and $F_3^{d/p}$ in LFQDM are similar to $F_3^{\pi}$
and $F_3^{K}$ in our model.

\subsection{PDF Limit}

\begin{figure*}[htbp]
	\centering
	\subfigure{\begin{minipage}[b]{0.4\linewidth}
			\centering
			\includegraphics[width=\linewidth]{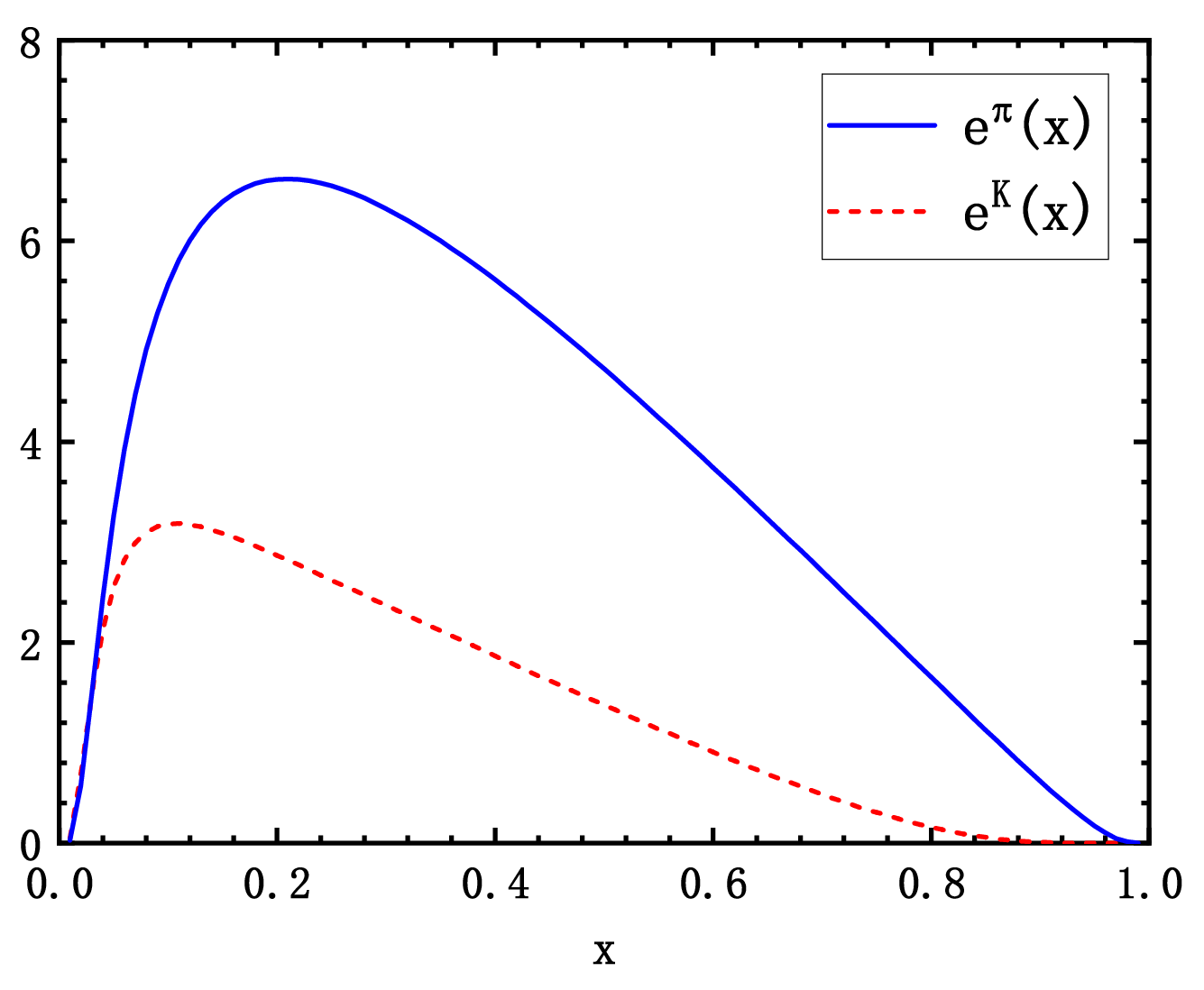}
	\end{minipage}}
	\subfigure{\begin{minipage}[b]{0.4\linewidth}
			\centering
			\includegraphics[width=\linewidth]{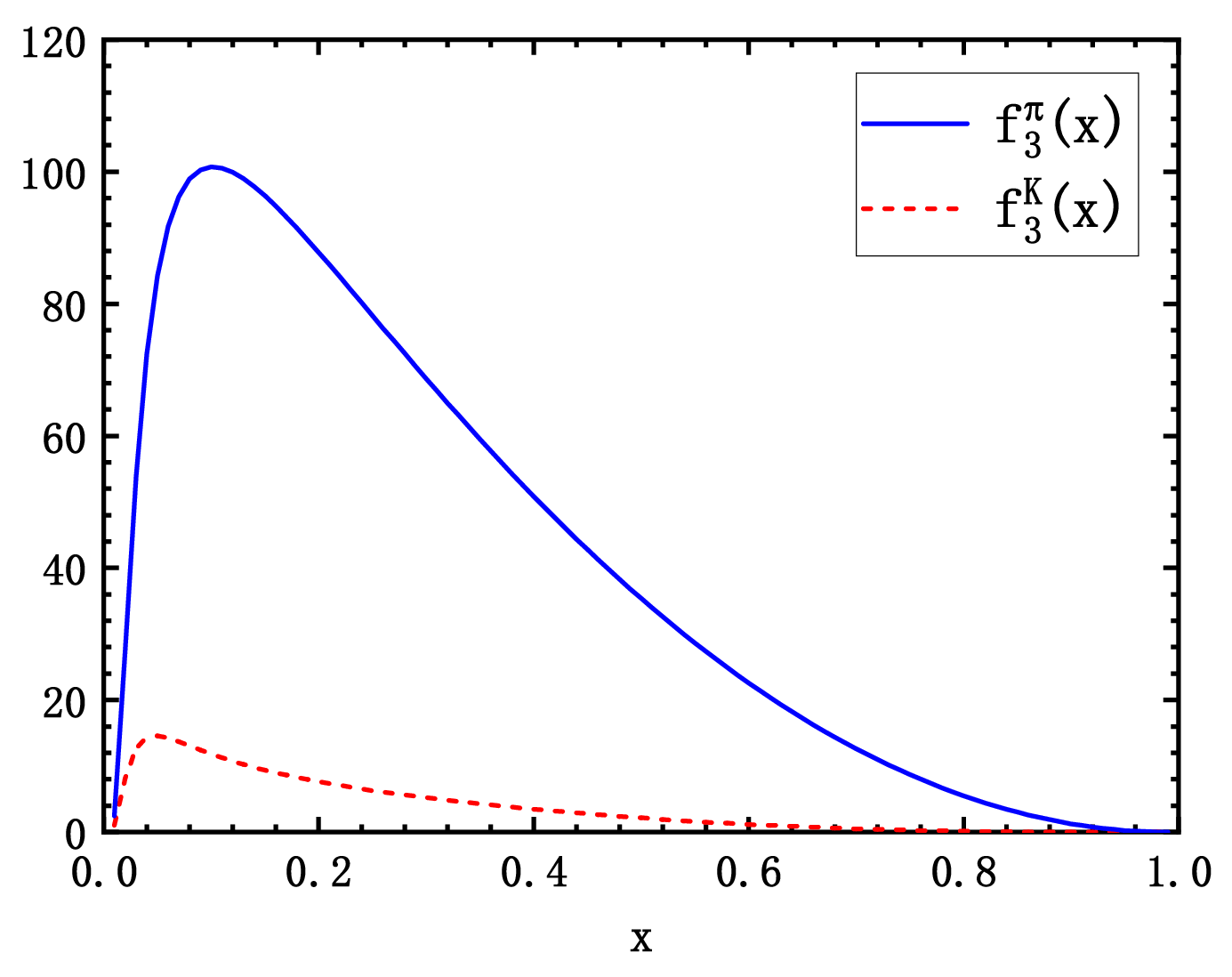}    	
	\end{minipage}}
	\caption{The twist-3 (left panel) and twist-4 (right panel) PDFs of the $u$ quark for the pion and kaon as functions of $x$ in the LCQM.} 
	\label{PDFs}      
\end{figure*}

For spin-0 pseudoscalar mesons, the PDFs are defined through the quark-quark correlation function~\cite{Jaffe:1991ra}
\begin{align}
	 \notag&{\Phi}^{[\Gamma]}(x)\\&= \int \frac{d z^{-}}{2 \pi}e^{ik \cdot z}\left. \left\langle p;S\left|\bar{\psi}\left(-\frac{z}{2}\right) \Gamma \psi\left(\frac{z}{2}\right)\right|p;S \right\rangle\right|_{z^{+} = 0, \vec{z}_{T} = 0},
\end{align}
All twist PDFs can be obtained by taking different cases of Dirac matrix structure $\Gamma$. Here we consider the case of $\Gamma=\gamma^{+}, 1, \gamma^{-}$
\begin{align}
    &{\Phi}^{[\gamma^{+}]}(x)= f_{1}(x) ,\\
	& {\Phi}^{[1]}(x)= \frac{M}{P^{+}}e(x),\\
    & {\Phi}^{[\gamma^{-}]}(x)=\frac{M^{2}}{(P^{+})^{2}} f_{3}(x).
\end{align}
Here, $f_{1}(x)$,  $e(x)$ and $f_{3}(x)$ are the twist-2, twist-3 and twist-4 PDFs, respectively.
Within the overlap representation in terms of LCWFs, the unpolarized PDF for the pion and kaon mesons can be calculated from
\begin{align}
	\notag f_{1}(x) & = \int \frac{d^{2} \boldsymbol{k}_{T}}{16 \pi^{3}}\left[\left|\psi_{0}\left(x, \boldsymbol{k}_{T}, \uparrow, \uparrow\right)\right|^{2}\right. \\\notag
	& +\left|\psi_{0}\left(x, \boldsymbol{k}_{T}, \uparrow, \downarrow\right)\right|^{2} \\\notag
	& +\left|\psi_{0}\left(x, \boldsymbol{k}_{T}, \downarrow, \uparrow\right)\right|^{2} \\
	& \left.+\left|\psi_{0}\left(x, \boldsymbol{k}_{T}, \downarrow, \downarrow\right)\right|^{2}\right].
\end{align}
Using the LCWFs in Eq.~(\ref{piLCWFs}) and Eq.~(\ref{KLCWFs}), we can obtain the valence quark unpolarized PDF of pion and kaon
\begin{align}
    f^{\pi}_{1}(x) & = \int \frac{d^{2} \boldsymbol{k}_{T}}{16 \pi^{3}}  |\varphi^{\pi}\left(x, \boldsymbol{k}_{T}\right)|^{2},\\
	\notag f^{K}_{1}(x) & = \int \frac{d^{2} \boldsymbol{k}_{T}}{16 \pi^{3}}\left[\left((1-x)m_{1}+xm_{2}\right)^{2}+ \boldsymbol{k}_{T}^{2}\right] 
	\\&\times\frac{|\varphi^{K}\left(x, \boldsymbol{k}_{T}\right)|^{2}}{\boldsymbol{k}_{T}^{2}+L^{2}}.
\end{align}
In the forward limit, the twist-3 GPD $E_{2}$ and twist-4 GPD $	F_{3}$ can reduce to their corresponding PDF 
\begin{align}
	E_{2}(x, 0,0) &  = e(x), \\
	F_{3}(x, 0,0) &  = f_{3}(x) .
\end{align}
We can easily verify that our results satisfy the following relations~\cite{Lorce:2016ugb}
\begin{align}
	 e(x) & = {1\over x}\frac{m_{q}}{M_{\pi(K)}} f_{1}(x), \label{eq:ef1}\\
    f_{3}(x) & ={1\over x^2} \int d^2 \bm{k} \frac{\boldsymbol{k}_{T}^{2}+m_{q}^{2}}{2M} f_{1}(x,\bm k_T^2),\label{eq:f3f1}
\end{align}
as well as the sum rule~\cite{Lorce:2016ugb}
\begin{align}
	\int d x x e^{q}(x) & = \frac{m_{q}}{M} N_{q},
\end{align}
where $N_{q}$ is the valence number of flavor $q$, specifically, $N_{u}=N_{\bar{d}}=1$ for $\pi^{+}$ and $N_{u}=N_{\bar{s}}=1$ for $K^{+}$.

In the left and right panels of Fig.~\ref{PDFs}, we depict the twist-3 PDF $e^{\pi(K)}(x)$ and the twist-4 PDF $f^{\pi(K)}_{3}(x)$ vs $x$, respectively. 
We find that both $e(x)$ and  $f_{3}(x)$ for pion and kaon are positive in the entire $x$ region.
The large size of  $e(x)$ and  $f_{3}(x)$ comes from the prefactors $1/x$ and $1/x^2$ in Eqs.~(\ref{eq:ef1}) and (\ref{eq:ef3f1}). 
Comparing the higher-twist PDFs for the pion and kaon, one can find that they have the similar shape but different size. Particularly, the size of $f_{3}^{\pi}(x)$ is much larger than that of $f_{3}^{K}(x)$.

Finally, we would like to compare our model results with the model calculations from other works~\cite{Lorce:2016ugb,Puhan:2023ekt,Zhu:2023lst}, where the higher-twist PDFs of pion and kaon have also been investigated. 
In Ref.~\cite{Lorce:2016ugb}, the LFCM was applied to calculate $e^{q}(x)$ and $f^{q}_{3}(x)$ of the pion. We find that our results for $e^{q}(x)$ and $f^{q}_{3}(x)$ agree with those in Ref.~\cite{Lorce:2016ugb}. $e^{q/\pi}(x)$ was also calculated in Ref.~\cite{Zhu:2023lst}, which showed a different tendency of $e^{q}(x)$ at small $x$ compared to both the LFCM result in Ref.~\cite{Lorce:2016ugb} and our results.  
At large $x$, $e^{q/\pi}(x)$ in the three models shares similar qualitative characteristics.
In Ref.~\cite{Puhan:2023ekt}, the light-cone quark model  and the light-front hadron model were utilized to calculate  $e^{q}(x)$, and $f^{q}_{3}(x)$ for both the pion and kaon, corroborating trends observed in our study. 
However, their investigation reveals that the peaks of twist-3 and twist-4 PDFs for pion are smaller than those for kaon, contrary to our results.

\section{CONCLUSION}\label{Sec5} 
 
In this work, we studied the twist-3 and twist-4 valence quark GPDs of the pion and kaon at zero skewness using a light-cone quark model, in which the higher-twist GPDs can be expressed as the overlaps of light-cone wave functions. 
The LCWFs of the spin-0 pseudoscalar mesons have been derived in terms of the $|q\bar{q} \rangle$ Fock states. 
Using the overlap representation of LCWFs, we presented the analytic expressions of the twist-3 GPDs $E^{\pi}_{2}(x,0,t)$, $G^{\pi}_{2}(x,0,t)$ and twist-4 GPDs $F^{\pi}_{3}(x,0,t)$, $H^{\pi}_{3}(x,0,t)$ of pion as well as the twist-3 GPDs $E^{K}_{2}(x,0,t)$, $G^{K}_{2}(x,0,t)$ and twist-4 GPDs $F^{K}_{3}(x,0,t)$, $H^{K}_{3}(x,0,t)$ of kaon. 
By properly choosing the values of the parameters in the model, we obtained the numerical results of the twist-3 and twist-4 GPDs of the pion and the kaon as functions of the longitudinal momentum fraction $x$ and the transverse momentum transfer ${\Delta}_{T}$. 
The relations between the twist-3 PDF $e(x)$ and twist-4 PDF $f_{3}(x)$ and unpolarized PDF $f_{1}(x)$ of the pion and kaon were verified by taking the forward limit. 
Furthermore, the numerical results for the corresponding twist-3 and twist-4 PDFs of the pion and kaon have also been shown in our work. 
We also compared our model results for $e^{q}(x)$ and $f^{q}_{3}(x)$ of the pion and the kaon to find some similarities and differences between them which need to be verified by future theoretical and experimental analysis. 
These results may provide useful information on the 3-D structure of the pion and the kaon.

\end{document}